\documentclass[manuscript,nonacm]{acmart}

\usepackage{cleveref}
\usepackage{booktabs}
\usepackage{tabularx}
\usepackage{array}
\usepackage{dirtytalk}
\usepackage{pdfpages}

\begin{document}

\title{XAI Evaluation Cards: A Practical Method for Designing Human-Centred XAI Evaluations}

%% Authors and affiliations (ACM acmart format).
%% Contribution roles previously in \credit (CAS) are not duplicated here;
%% add an Author Contributions section in the manuscript if your venue requires it.

\author{Kristýna Sirka Kacafírková}
\orcid{0000-0002-5910-4408}
\affiliation{%
  \institution{imec-SMIT, Vrije Universiteit Brussel}
  \city{Brussels}
  \country{Belgium}}
  \email{kristyna.kacafirkova@vub.be}
\authornote{Corresponding author.}
  
  \author{Ivania Donoso-Guzmán}
\orcid{0000-0002-2427-9128}
\affiliation{%
  \institution{KU Leuven, Department of Computer Science}
  \city{Leuven}
  \postcode{B-3001}
  \country{Belgium}}
\affiliation{%
  \institution{Augment, imec research group at KU Leuven}
  \city{Leuven}
  \postcode{B-3001}
  \country{Belgium}}
\affiliation{%
  \institution{Computer Science Department, Pontificia Universidad Católica de Chile}
  \city{Santiago}
  \country{Chile}}

\author{Denis Parra}
\orcid{0000-0001-9878-8761}
\affiliation{%
  \institution{Computer Science Department, Pontificia Universidad Católica de Chile}
  \city{Santiago}
  \country{Chile}}

\author{Katrien Verbert}
\orcid{0000-0001-6699-7710}
\affiliation{%
  \institution{KU Leuven, Department of Computer Science}
  \city{Leuven}
  \postcode{B-3001}
  \country{Belgium}}
\affiliation{%
  \institution{Augment, imec research group at KU Leuven}
  \city{Leuven}
  \postcode{B-3001}
  \country{Belgium}}
  
  \author{An Jacobs}
\orcid{0000-0001-8058-1455}
\affiliation{%
  \institution{imec-SMIT, Vrije Universiteit Brussel}
  \city{Brussels}
  \country{Belgium}}

\renewcommand{\shortauthors}{Sirka Kacafírková et al.}

\begin{abstract}
Evaluating explainable AI (XAI) systems from a human-centred approach requires researchers to select from numerous evaluation dimensions and measures, often in an ad hoc and fragmented manner. This paper introduces a method to help HCI, computer science, designers and social science researchers systematically evaluate XAI systems. The approach is based on an updated XAI-specific evaluation framework derived from an analysis of 82 studies. Using this framework, we developed a card-sorting method with 36 cards to help researchers prioritise relevant evaluation aspects. The process was tested with two research groups (n = 13) across five projects. The XAI Evaluation Cards are available as a printable appendix, along with an online repository of methods from previous XAI studies. Although not exhaustive, our findings indicate that the card-sorting approach can organise and streamline the design of the evaluation process, encouraging a more comprehensive and multidisciplinary assessment of XAI systems in research and development.
\end{abstract}



\begin{teaserfigure}
    \Description {Three-step overview of the XAI Evaluation Cards method. Step 1, identifying relevant properties: research teams reflect on their use case and sort evaluation-property cards into relevant and not relevant categories. The cards represent five main categories: Objective System Aspects, User Experience, Subjective System Aspects, Explanation Aspects, and Interaction. Step 2, prioritisation: the cards considered relevant are discussed and ranked according to importance, with rank 1 representing the highest priority. Example ranked properties shown are Confidence, Cognitive Load, and Representativeness. Step 3, finding measurements: teams select the properties they intend to evaluate and consult the accompanying online repository to explore how these properties have been operationalised and measured in existing studies.}
 \includegraphics[width=\textwidth]{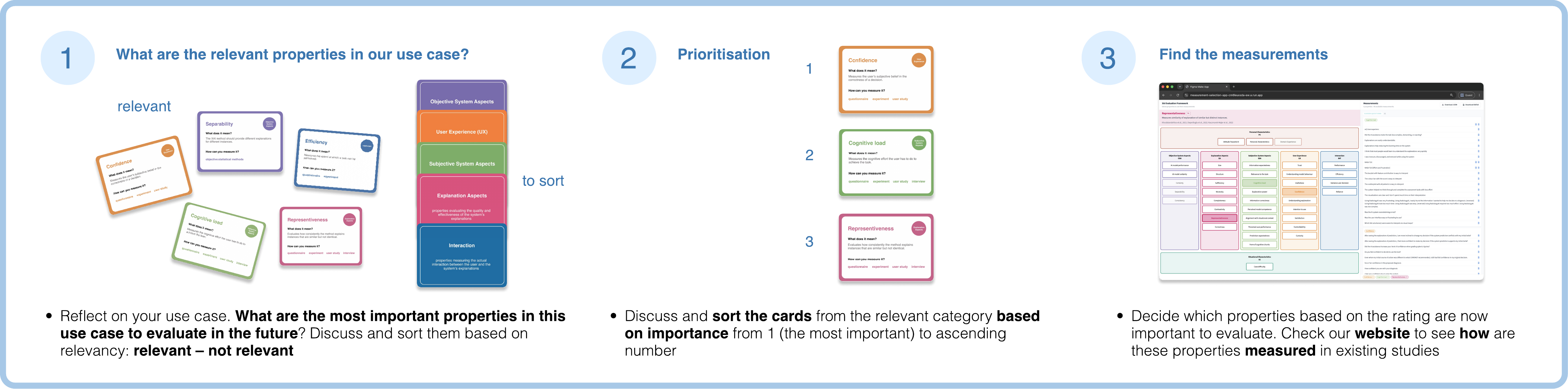}
  \caption{3-step guide on how to use the XAI evaluation cards; The printable version of the cards is available in Appendix A; The website of the repository of measurements for each property is available \href{https://measurement-selection-app-cm6leuxsda-ew.a.run.app/}{online}.}
  \label{fig:teaser}
\end{teaserfigure}

\maketitle

\section{Introduction}

Evaluation is a fundamental component of the development of explainable AI (XAI) systems. It enables researchers to assess whether explanations achieve their intended purpose, examine relationships between different evaluation aspects, and identify inconsistencies across previous findings. A systematic approach to evaluation can therefore contribute to more transparent, comparable, and comprehensive XAI research.

XAI systems can be evaluated both with and without users, and these approaches provide different but complementary forms of evidence.  Technical evaluation can assess properties such as model performance, fidelity, or robustness, whereas human-centred evaluation is needed to understand how explanations are perceived, interpreted, and used in practice. Although subjective measures have sometimes been criticised for their dependence on users' perceptions and context \cite{Zhou2021EvaluatingMetrics}, relying exclusively on objective measures provides only a partial account of explanation quality. Combining objective and subjective approaches can therefore support a more comprehensive evaluation, particularly within interdisciplinary teams \cite{Bennett2026, Schmid2022WhatFar, Mohseni2021ASystems}.

A growing body of literature has attempted to collect, categorise, and structure XAI dimensions that can be measured (such as \cite{Mohseni2021ASystems, Lopes2022XAIMethods, Rong2024TowardsExplanations, Vilone2021NotionsIntelligence, Lofstrom2022AMethods,Nauta2023FromAI}).
Nevertheless, existing resources often emphasise technically oriented approaches aimed primarily at developers and data scientists (e.g. \cite{Nauta2023FromAI}), while providing less support for selecting human-centred evaluation properties and measures. Moreover, reviews and frameworks frequently identify \textit{what} could be evaluated (or what has been currently evaluated) without providing sufficient practical guidance on \textit{how} research teams should select, prioritise, and operationalise these properties for a specific use case \cite{10.1145/3769678}. The absence of consistent terminology and unified classifications further complicates this process. Closely related concepts may be described using different terms across disciplines, while concepts such as explainability and interpretability themselves continue to be used inconsistently \cite{Graziani2023ASciences}.

As a result, XAI evaluation remains fragmented and often conducted ad hoc, while empirical evaluations involving users remain comparatively limited \cite{Suh2025FewerHumans}. Selecting appropriate evaluation properties and measures can become a lengthy and uncertain process. Researchers may rely on familiar, frequently cited, or readily available measures rather than those best suited to their particular system, users, and design stage of development (for example, the widely used measurements by \citet{Hoffman2018MetricsCognition}). This challenge becomes especially prominent in interdisciplinary XAI projects, where computer scientists, social scientists, designers, and domain experts may approach evaluation from different conceptual and methodological traditions. Planning of the evaluation design therefore involves not only identifying suitable measures, but also establishing a \textit{shared understanding} of what is relevant to evaluate, what should be prioritised at a particular design stage of the project, and how these properties can subsequently be operationalised as an iterative and continuous process in all design stages.  

Although recent work has provided greater conceptual and methodological structure for XAI evaluation, the translation of this knowledge into practical methods for evaluation planning remains comparatively limited. Other areas of XAI have already produced more actionable resources to support design activities. For example, the XAI Question Bank developed by \citet{Liao2020QuestioningExperiences} provides practitioners with a structured method to identify and design appropriate user-centred explanations. Comparable practical support for deciding what should be evaluated from human centered approach, what should be prioritised in a particular context, and how these decisions can be made collaboratively within interdisciplinary teams remains less developed. To address this gap, we introduce the XAI Evaluation Cards, an evidence-informed method designed to support interdisciplinary research teams in structuring and prioritising their evaluation planning across different stages of project development. The method helps teams identify potentially relevant evaluation properties, discuss and prioritise them in relation to their specific use case and project stage, and subsequently explore suitable ways of measuring them.

The main contribution of this paper is therefore the XAI Evaluation Cards method, together with insights from its initial development and evaluation. The cards translate an evaluation framework derived from an analysis of 82 XAI studies into a structured conversation and prioritisation tool. Combined with an accompanying repository of previously used evaluation measures, the method supports a three-step process: (1) identifying what may be relevant to evaluate, (2) prioritising what should be evaluated in the current context and design stage, and (3) exploring how the selected properties can be operationalised. This pilot study provides an initial evaluation of how the cards support the identification and prioritisation processes that are essential to be able to operationalise the evaluation.

\section{Related work}

\subsection{The Landscape of XAI Evaluation Frameworks}

Several surveys have attempted to structure the XAI evaluation landscape by interpreting primary evidence. Early foundational works established general guidelines for multidisciplinary design and evaluation \citep{Mohseni2021ASystems, Vilone2021NotionsIntelligence}, while more recent efforts have moved toward objective, computational metrics \citep{Nauta2023FromAI}.

In parallel, a human-centred pivot has emerged. Surveys by \citet{Lopes2022XAIMethods} and \citet{Nguyen2024HowSurvey} provide maps for systematically selecting user-centred strategies during the development process. Furthermore, systematic reviews by \citet{Kim2024Human-centeredReview} and \citet{Rong2024TowardsExplanations} categorise meaningful explanations into dimensions of trust, usability, and human-AI performance. Despite this progress, the field remains fragmented; many studies do not follow a formal framework \citep{Lopes2022XAIMethods} or continue to evaluate systems without human participants \citep{Suh2025FewerHumans}.

\begin{figure}[!ht]
    \centering
    \Description{Matrix comparing concepts defined in seven XAI evaluation frameworks: Jung et al. (2023), Nauta et al. (2023), Lopes et al. (2022), Vilone and Longo (2021), Löfström et al. (2022), Mohseni et al. (2021), and Rong et al. (2024). Each column represents one framework, while vertically aligned boxes indicate concepts with overlapping definitions across frameworks. The figure shows substantial variation in terminology and level of granularity. Similar ideas are described using different terms, for example around satisfaction, interpretability, comprehensibility, mental models, trust, and performance, while some broader concepts span several more specific concepts used in other frameworks. Horizontal lines separate conceptual clusters, and dotted boundaries indicate concepts that extend across clusters. Overall, the figure illustrates the lack of one-to-one alignment between XAI evaluation concepts and highlights considerable conceptual overlap across existing frameworks. }
    \includegraphics[width=\linewidth]{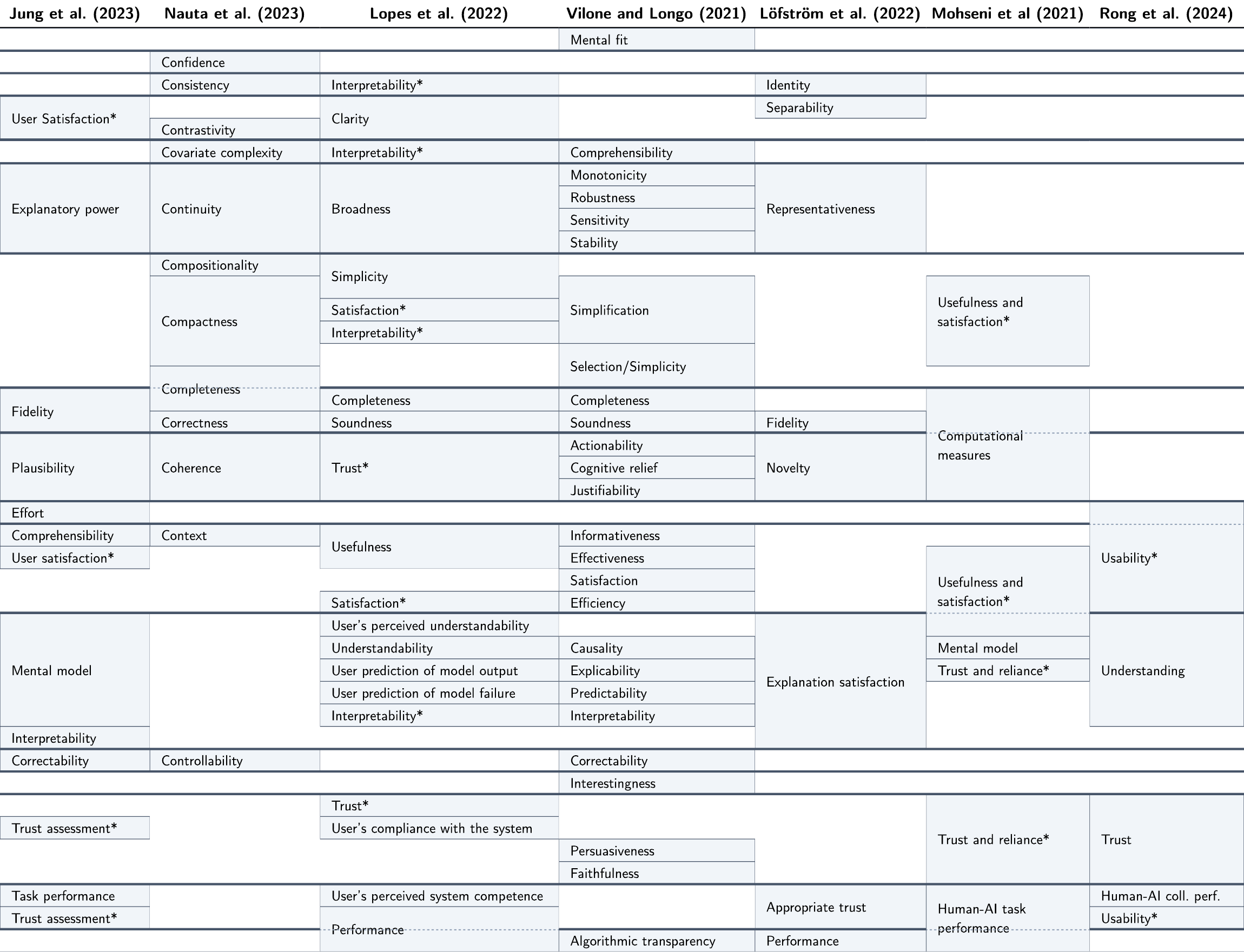}
    \caption{Concepts defined by XAI evaluation frameworks. Each study is a column. Each row represents a concept definition from these studies. Concepts spanning across multiple rows are conflating with concepts from other studies. Concepts marked with \ding{83} span multiple non-continuous rows. Solid lines separate distinct clusters of concepts. Concepts crossed by dotted lines span over different clusters. These frameworks were selected because they propose a set of concepts for evaluating XAI experiences through analysis of empirical studies in the field.}
    \label{fig:related_work:XAI_evaluation_frameworks}
\end{figure}

A significant barrier to progress is the lack of standardisation. Most existing frameworks propose unique definitions of properties such as trust or satisfaction rather than building on prior work, while others, like \citet{Kim2024Human-centeredReview} or \citet{Nguyen2024HowSurvey}, do not provide the definitions of the concepts they identify. \Cref{fig:related_work:XAI_evaluation_frameworks} displays the overlapping concepts among the 7 identified evaluation frameworks in the area that explicitly define concepts to be evaluated based on empirical evidence. Each study is represented by a column, and the properties they defined appear as rows. Concepts spanning across multiple rows are conflating with concepts from the other studies. For instance, \textit{User Satisfaction} from \citet{Jung2023EssentialReview}, defined as \say{Identifies the explanations' \textit{clarity} and utility}, contains the meaning of \textit{Clarity} from \citet{Lopes2022XAIMethods}, defined as \say{implies that the explanation should be unambiguous}. \textit{Contrastivity} from \citet{Nauta2023FromAI}, \say{Describes how discriminative the explanation is w.r.t. other events or targets}, and \textit{Separability} from \citet{Lofstrom2022AMethods}, \say{Non-identical instances should not have identical explanations}, overlap with the notion of \textit{clarity} of these two previous concepts but do not overlap with each other.
Frameworks proposed by \citet{Mohseni2021ASystems}, \citet{Lopes2022XAIMethods}, and \citet{Rong2024TowardsExplanations} tend to employ broader categories that cluster diverse measurements. In contrast, the approaches of \citet{Vilone2021NotionsIntelligence}, \citet{Lofstrom2022AMethods}, and \citet{Nauta2023FromAI} provide more granular terms that aim to isolate factors that influence the systems' outcomes. 

A similar phenomenon occurs with the measurements to evaluate \textit{Trust} and \textit{Performance}. While \citet{Jung2023EssentialReview} and \citet{Rong2024TowardsExplanations} assess \textit{Trust} through the behavioural proxy of agreement between the user and the AI model, \citet{Mohseni2021ASystems} utilise similar agreement-based measures to evaluate the human-AI \textit{Performance}. This highlights a significant overlap where the same behavioural data is used to justify fundamentally different constructs. 

These conceptual inconsistencies arguably pose a risk to the scientific rigour of the XAI field. As \citet{Clinciu2019ATerminology} suggest, the lack of a shared vocabulary and clear definitions makes it difficult to progress the field. Particularly in XAI, it becomes difficult to identify what effects explanations have on users. These inconsistencies may lead to a fragmented research landscape in which different studies appear to yield contradictory results, yet these discrepancies might stem from researchers unintentionally measuring different phenomena under the same label. 

To summarise, existing literature provides a ``menu'' of what to evaluate, but leaves a substantial gap in guiding teams on how to select and operationalise these properties for specific contexts.

\subsection{Navigating interdisciplinary collaboration in XAI evaluation}\label{sec:rw:interdisciplinary}

The conceptual inconsistencies in XAI frameworks are further exacerbated by the practical realities of system development. AI products are typically built by cross-functional teams, yet HCI research consistently finds that AI work disrupts traditional design and engineering workflows \citep{Yang2020Re-examining, Piorkowski2021How, Zdanowska2022A}. Specifically, human-AI systems resist the traditional ``separation of concerns'' because user needs must shape not only the interface but also core model components and training data \citep{Subramonyam2022Solving, Muralikumar2025An}. As \citet{Yang2020Re-examining} establish, designing human-AI interaction is uniquely difficult because designers struggle to grasp the boundaries of AI capabilities, while engineering workflows often optimise capabilities without grounding them in concrete user needs \citep{Subramonyam2022Solving, Yildirim2023Investigating}.

Consequently, UX and AI work often proceeds independently and linearly, with user-centred design entering the process only after core model decisions have already been made \citep{Feng2023How, Subramonyam2021Towards, Bruun2025Coordination}. This late convergence pushes human-centred evaluation into sequential downstream work, despite the fact that AI uncertainty requires continuous, iterative coordination \citep{Bruun2025Coordination, Muralikumar2024Analyzing}. The demand for Responsible AI further sharpens this divide, as UX practitioners are frequently required to perform extra ``bridging labour'' to translate ethics and user feedback into formats actionable for data-centric engineering cultures \citep{Wang2023Designing, Varanasi2023It, Sambasivan2021Everyone}.

A fundamental driver of this ML--UX divide is the lack of a shared language for negotiating AI system requirements and evaluation metrics \citep{Yildirim2023Investigating, Moore2023fAIlureNotes, Wang2023Designing}. This is not merely generic miscommunication; it reflects role-specific definitions of key terms, divergent success criteria, and uneven visibility into model development \citep{Muralikumar2024Analyzing, Piorkowski2021How}. For example, \citet{Almahmoud2021How} demonstrate that data scientists typically discuss system ``performance'' in terms of model accuracy or recall, whereas customer-facing stakeholders evaluate ``performance'' based on overall user task success. When teams use the same terms differently, or different terms for the same underlying concern, vocabulary itself becomes a major coordination barrier \citep{Yildirim2023Investigating, Feng2023How}. Without comprehensible, contextual model information, specifications become notoriously difficult to translate into shared design and evaluation decisions.

To overcome this structural mismatch between model-centric and user-centric evaluation, recent HCI literature highlights the need for explicit translation infrastructure. Researchers have begun exploring how shared artefacts can puncture role boundaries and help teams negotiate requirements together. For instance, \citet{Liao2020QuestioningExperiences} introduced an XAI question bank that reframes explainability around the concrete questions users actually ask, explicitly bridging the gap between algorithmic outputs and user needs. Similarly, \citet{Subramonyam2022Solving} highlight how practitioners invent ad hoc ``leaky abstractions'' to share low-level AI constraints across expertise lines, while tools like \textit{fAIlureNotes} \citep{Moore2023fAIlureNotes} and tailored onboarding materials \citep{Cai2021Onboarding} help make model limitations discussable across roles.

Ultimately, the core challenge for interdisciplinary XAI teams is less about interpersonal conflict and more about the lack of robust translation tools connecting disparate evaluation priorities. Designing a successful evaluation of an XAI system requires an explicit method for making these differing interpretations visible, establishing a shared vocabulary, and negotiating what should be prioritised at a given stage of development.

\subsection{Card-based methods and XAI domain}

To bridge the interdisciplinary divides described above, HCI research frequently relies on \say{Boundary Objects} \cite{STAR198937}, artefacts that are flexible enough to adapt to the specific needs of different disciplines, yet robust enough to maintain a common identity across them. Card-based methods are a well-established medium for this in design research. Tangible cards prompt reflection, introduce alternative perspectives, and structure collaborative decision-making by making abstract concepts visible and physically manipulable \cite{Aarts2020DesignSets}. Rather than simply being used to uncover users' mental models or organise information \cite{Tchivi2025ASorting, Macias2021EnhancingAnalysis}, cards support collective sense-making. They help interdisciplinary groups negotiate shared priorities by providing a constrained, shared vocabulary that levels the playing field between technical and non-technical stakeholders \cite{Sandnes2026NoDecision-Making}.

In the XAI domain, card-based methods have recently emerged to support system development. Previous studies have used classical card-sorting techniques to elicit users' needs and preferences regarding explainability \cite{Morandini2025UserStudy}. More recent work has introduced design-card approaches to support the design of explanations in AI systems \cite{Naiseh2026DesigningCards, Lee2024TheExplanations}.

However, while these existing XAI card toolkits are highly valuable, they are primarily \textit{generative}, they focus on the creation of explanations and the elicitation of user needs. There remains a lack of tools designed to support the \textit{evaluative} phase of XAI development in multidisciplinary teams. Consequently, teams lack collaborative methods to navigate the complex, fragmented evaluation frameworks discussed in \Cref{sec:rw:interdisciplinary}, leaving them without a structured way to plan the actual scientific measurement and testing of their systems.

Our work also builds on research that examines how XAI evaluation criteria are prioritised. Liao et al. \cite{Liao2022ConnectingAI}, for example, assessed the perceived importance of different criteria for XAI evaluation through survey-based quantitative ratings. While this work provides vital insight into which evaluation criteria the community generally considers important, our approach treats these judgments as explicitly context-dependent. Rather than establishing a fixed hierarchy of evaluation criteria, card-based boundary objects invite research teams to determine which properties are relevant for their particular use case and current stage of development, and to negotiate their priorities collectively. This acknowledges that an evaluation property considered critical in one context may be entirely irrelevant, or not yet appropriate to assess, in another. The aim is therefore not to prescribe a universal set of evaluation rules, but to support teams in reaching a situated and shared evaluation plan.

Drawing on boundary objects, design cards, and efforts to operationalise existing evaluation frameworks, we developed an adapted card-based method for interdisciplinary XAI evaluation planning. Rather than asking participants to group cards into conceptual categories, the method invites teams to distinguish between relevant and non-relevant evaluation properties and subsequently prioritise those considered most important for their specific project and design stage. In doing so, the exercise makes the broader evaluation space visible while supporting teams in translating abstract metrics into a concrete action plan.

In the following sections, we describe the development of the XAI Evaluation Cards and their accompanying measurements repository (\Cref{sec:cards_dev}), followed by an empirical user study testing the method with cross-functional research teams (\Cref{sec:study}).

\section{Development of the XAI Evaluation Cards Method} \label{sec:cards_dev}
The design of the XAI Evaluation Cards was informed by a two-stage study. Stage 1 consisted of a systematic review of 82 XAI studies and the iterative development of an evaluation framework to identify how XAI systems have been evaluated and which properties and measures have been used. In Stage 2, this framework was translated into the XAI Evaluation Cards. In the following sections, we describe how the findings from Stage 1 informed the development of the XAI Evaluation Cards (the main contribution of this paper)  and detail its intended usage.

\subsection{Stage 1: Systematic review of evaluation in existing XAI studies}
Following a validated search strategy across five academic databases (Scopus, Web of Science, ACM, IEEE, and PubMed), we identified 13,860 records. After duplicate removal, 8,226 unique records proceeded to a multi-stage screening process. Studies were included if they were peer-reviewed, involved a functional AI implementation with algorithmically generated explanations, and reported an empirical user evaluation in a medically relevant application. This process resulted in a final dataset of 82 papers.

Stage 1 consisted of study identification and selection with the iterative development of a taxonomy for analysing XAI evaluations. This last process resulted in a framework that was conceptually grounded in existing literature and empirically informed by healthcare XAI research. We then applied the framework to the 82 included studies to examine how different dimensions of XAI evaluation were operationalised and which properties and measures were used. The aim was to identify recurring patterns that could inform decisions about what to evaluate in XAI studies. Three HCI researchers (two with computer science backgrounds and one with a sociology background) coded the studies concurrently and met weekly over a six-month period to resolve ambiguous cases, assess conceptual saturation, and refine the framework through consensus.

Based on this analysis, we built upon the existing XAI evaluation framework originally proposed by \citet{Donoso-Guzman2023TowardsAI}. Specifically, we refined and updated the definitions of its properties and dimensions to better reflect empirical practices. A detailed account of the systematic review that produced these updates, including the coding process and resulting framework, is provided in our previous study \footnote{Due to the anonymisation policy, we cannot refer to the paper directly in the manuscript. We have therefore included it as supplementary material to provide reviewers with access to additional methodological details if needed.}.

\begin{figure}[!ht]
    \centering
    \Description{Framework showing XAI evaluation properties organised by conceptual component. The framework contains seven categories: Objective System Aspects (OSA), Explanation Aspects (EA), Subjective System Aspects (SSA), User Experience (UX), Interaction, Personal Characteristics (PC), and Situational Characteristics (SC). Following Knijnenburg and Willemsen (2015), Objective System Aspects influence Subjective System Aspects, which in turn affect User Experience and can modify Interaction. These components are also influenced by Personal and Situational Characteristics. Donoso-Guzmán et al. (2023) extended this structure by adding Explanation Aspects, which capture the quality of the selected reasons underlying an explanation and are proposed to influence Subjective System Aspects.

Objective System Aspects include AI model performance, AI model certainty, certainty, consistency, and separability. Explanation Aspects include necessity, sufficiency, correctness, completeness, contrastivity, size, structure, and representativeness. Subjective System Aspects include alignment with situational context, cognitive load, explanation power, form of cognitive chunks, information correctness, information expectedness, perceived model competence, perceived user performance, prediction expectedness, and relevance to the task. User Experience includes confidence, controllability, curiosity, intention to use, satisfaction, trust, understanding the model, understanding the explanation, and usefulness. Interaction includes efficiency, performance, reliance, and variance in user decision. Personal Characteristics include domain experience and attitude towards AI, while Situational Characteristics include case difficulty.

Properties newly identified in this study are marked with an asterisk. These are domain experience, attitude towards AI, representativeness, information correctness, perceived user performance, prediction expectedness, confidence, intention to use, understanding the model, understanding the explanation, variance in user decision, and case difficulty.}
    \includegraphics[width=\linewidth]{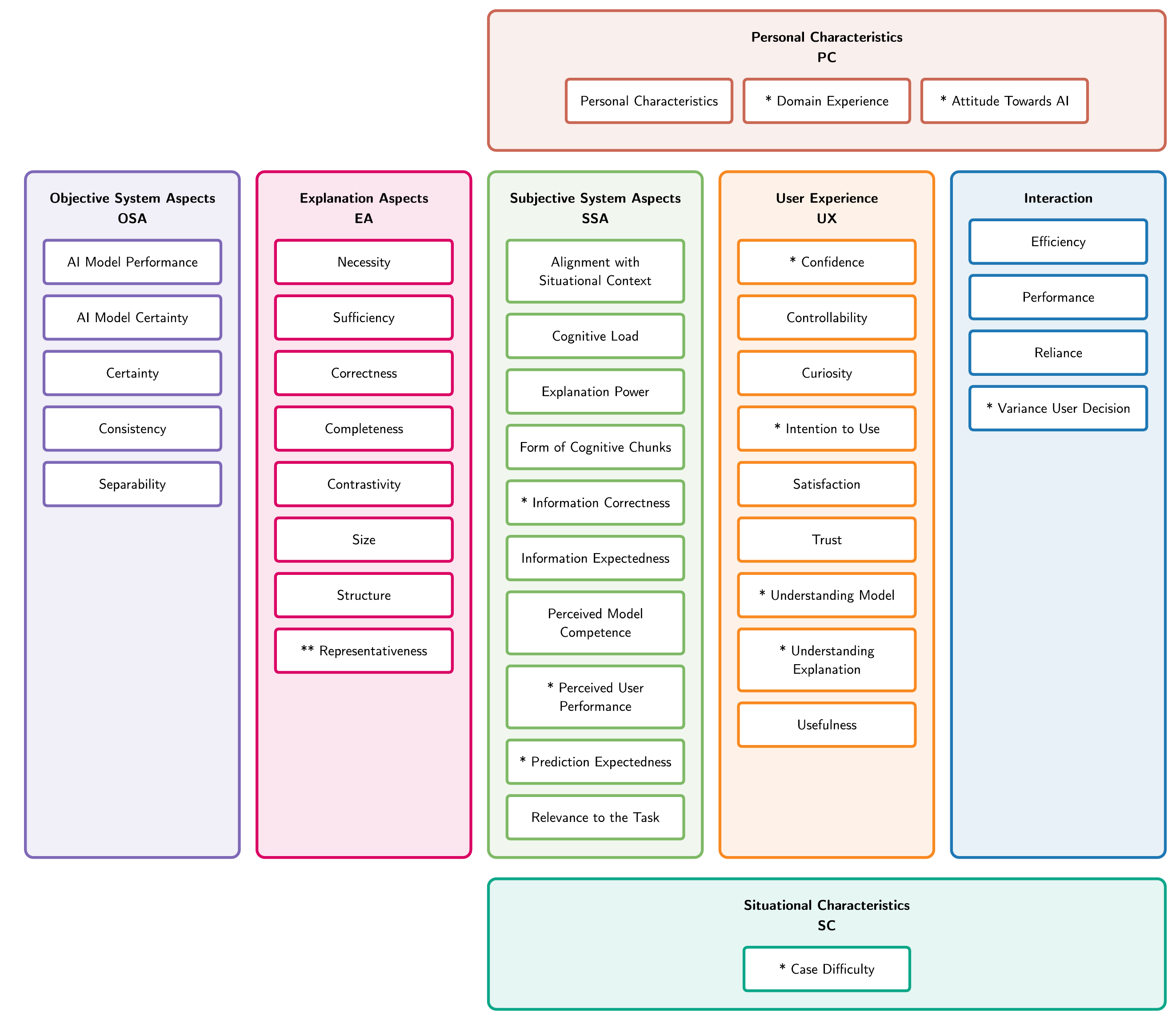}
    \caption{Framework properties organised by conceptual component. As stated by \citet{Knijnenburg2015EvaluatingExperiments}, \textit{Objective system aspects} influence the \textit{Subjective system aspects}, which in turn affect the \textit{User experience} and can modify the \textit{interaction}. These three components are influenced by \textit{Personal} and 
    \textit{Situational Characteristics}. \citet{Donoso-Guzman2023TowardsAI} added the \textit{Explanation aspects} to account for the quality of the selected reasons of the explanation and proposed that it influences \textit{Subjective system aspects}. New properties, identified in this study, are marked with \ding{83}. The definitions and associated papers can be accessed at \href{https://measurement-selection-app-cm6leuxsda-ew.a.run.app/}{the online website}.}
    \label{fig:Framework}
\end{figure}

The framework of properties for XAI evaluation updates the previous definitions from \cite{Donoso-Guzman2023TowardsAI} while retaining the same structure of 7 conceptual components, as shown in \Cref{fig:Framework}.
The \textit{Objective system aspects} represent technical features of the AI model and XAI method, such as \textit{AI model performance} and \textit{Certainty}, while \textit{Explanation aspects} assess the quality of the generated explanation product, including \textit{Correctness}, \textit{Completeness}, and \textit{Size} of the explanation. The \textit{Subjective system aspects} capture users' perceptions of these objective features, such as \textit{Explanation power}, \textit{Cognitive load}, and \textit{Relevance to the task}. 

\textit{User experience aspects} evaluate psychological outcomes, such as \textit{Trust}, \textit{Understanding model behaviour}, and \textit{Satisfaction}, whereas \textit{Interaction} measures behavioural outcomes and adoption potential, including \textit{Efficiency} and \textit{Reliance}. Finally, user and environmental variations are captured by \textit{Personal characteristics}, which account for individual dimensions that can affect the interaction (e.g., \textit{Domain experience}), and \textit{Situational characteristics}, which account for contextual factors beyond the system's control, such as \textit{Case difficulty}.

\subsection{Stage 2: Translation of the framework into cards}
\begin{figure}[hbt!]
    \centering
    \Description{Example of the front and back of an XAI Evaluation Card, connected by a two-way arrow to indicate that they are two sides of the same card. The front presents the evaluation property “Understanding Explanation,” which belongs to the User Experience category. It includes a definition stating that the property measures the user’s ability to correctly interpret the explanation generated by the system. It also lists example ways in which the property can be measured: questionnaire, experiment, user study, and interview. The back of the card identifies the broader category, “User Experience (UX),” and defines it as properties assessing what users perceive, feel, and experience when interacting with the system.}
    \includegraphics[width=\linewidth]{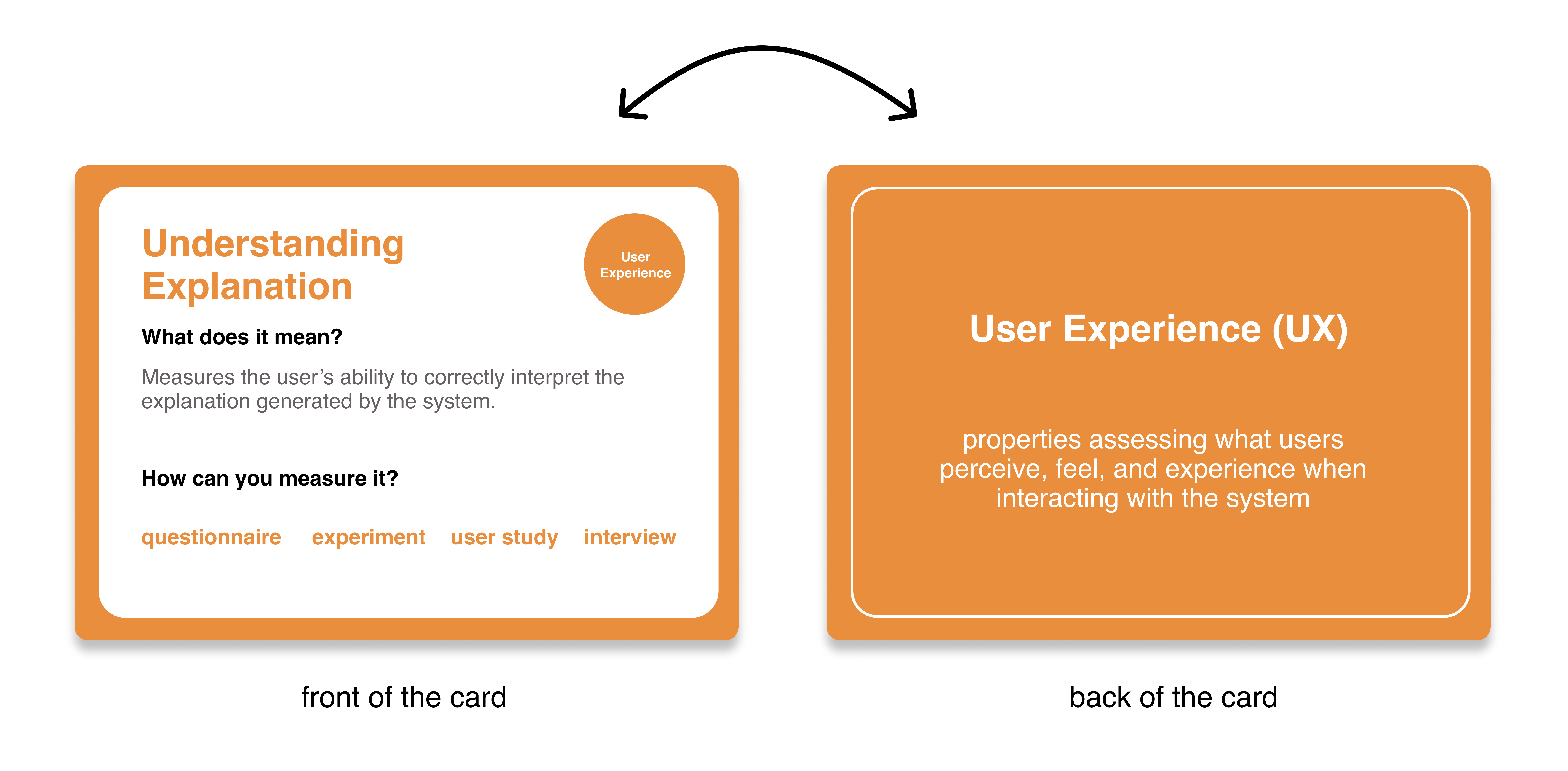}
    \caption{Example of the design and structure of an XAI Evaluation Card}
    \label{fig:Card example}
\end{figure}

Throughout the review, we observed limited reporting of the evaluation planning process. In most studies, the choice of what to evaluate was not explicitly justified or linked to a structured or reflective planning approach. Furthermore, quantitative methods were most common (n = 41), followed by mixed methods (n = 31) and qualitative methods (n = 10). The methodological approach was also associated with the number of properties evaluated. Quantitative studies assessed approximately four properties on average, qualitative studies close to five, and mixed-methods studies almost seven. This pattern might suggest that the range of properties considered is partly shaped by methodological choices and the availability of established measures, rather than being driven solely by their contextual relevance to the specific use case. The properties evaluated most frequently were also concentrated around a relatively small number of concepts, such as understanding and trust. However, our framework indicates that the evaluation space is considerably broader. These observations further support the need for a more reflective and structured approach to evaluation planning, in which the broader evaluation space is made visible before research teams narrow it down to the properties most relevant to their specific use case and design stage, and subsequently determine how these properties should be measured.

To operationalise this approach, make the framework more actionable, and support the selection of relevant evaluation properties, we translated the properties identified in Stage 1 into a set of XAI Evaluation Cards (see \Cref{fig:Card example}). Each property definition from the framework was reviewed for clarity and accessibility. Definitions that were considered sufficiently clear were retained, while others were revised with the support of ChatGPT (GPT-5.1 family) \footnote {Prompts used: reformulate this in more understandable sentence; check this definition and adjust if necessary; make it understandable to lay users as well; make it accessible}, which was used to suggest simpler formulations for readers who might be unfamiliar with the underlying terminology. All suggested revisions were subsequently reviewed by the first author before inclusion in the cards.

We focus on five main categories (see \Cref{fig:Framework}): \textit{Objective System Aspects}, \textit{Explanation Aspects}, \textit{Subjective System Aspects, User Experience, and Interaction.} We did not include \textit{Personal Characteristics }and \textit{Situational Characteristics} in the card set. 
Although these dimensions are important when designing an evaluation, they function primarily as contextual or individual-level control variables rather than as properties of explainability itself. Nevertheless, they should always be considered in relation to the specific use case and the users involved.

 For each property, we revisited the Stage 1 corpus to identify how it had been operationalised in previous XAI studies. Based on these examples, we added a brief \textbf{“How to measure it”} section to each card. The cards are intended to support research teams in discussing and prioritising the properties most relevant to their specific use case, while also serving as a structured conversation tool during evaluation planning.

Each card, therefore, contains the category to which the property belongs, a concise definition of the property, and examples of measurement approaches previously used to evaluate it (see \Cref{fig:Card example}).  These examples are intended to illustrate possible ways of operationalising a property rather than to prescribe a particular measurement method.

\begin{figure}[!ht]
    \centering
    \Description{Example printable sheet of XAI Evaluation Cards arranged in a three-by-three grid. The upper six cards belong to the Objective System Aspects category and are shown in purple. They represent AI Model Performance, Consistency, Separability, AI Model Certainty, Certainty, and Continuity. Each card contains the property name, a short definition, and a section indicating how the property can be measured. For these objective properties, the suggested approach is objective statistical methods. The lower three cards belong to the User Experience category and are shown in orange. They represent Trust, Usefulness, and Understanding Model. Each includes a short definition and examples of possible measurement approaches, including questionnaire, experiment, user study, and interview. The page includes cutting guides and a scissors symbol to indicate that the cards are intended to be printed and cut out for use during the card-sorting activity.}
    \includegraphics[width=\linewidth]{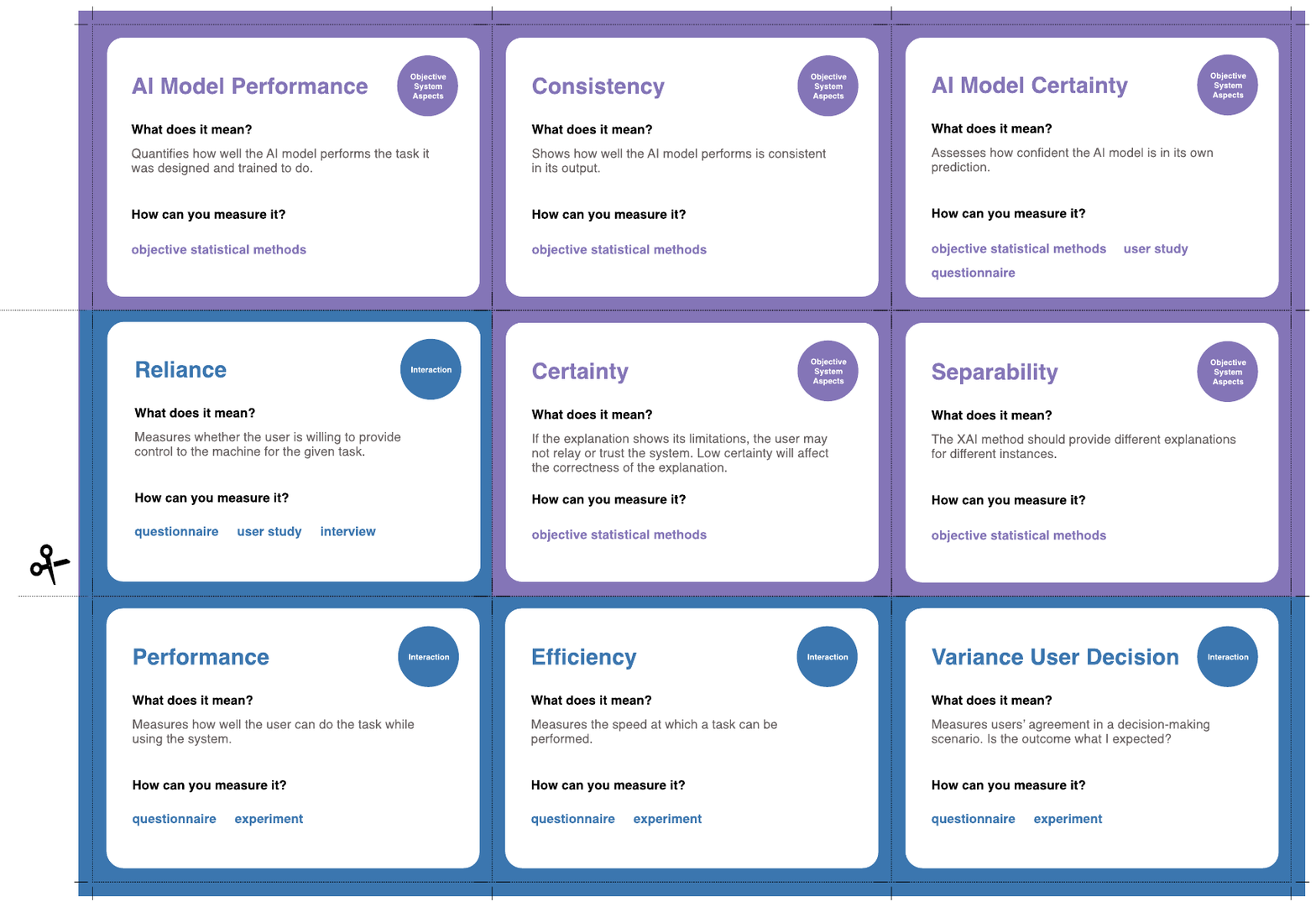}
    \caption{Example of printable XAI Evaluation Cards. The full printable template is provided in Appendix A, beginning on page 23.}
    \label{fig:Printable cards}
\end{figure}

We used a standard poker-card format with the cards arranged in an A4 template for printing and use in face-to-face sessions (see \Cref{fig:Printable cards}). The cards were manually designed by the authors using Figma (version 126.8.18). We also included blank cards for each category, allowing research teams to add properties that are relevant to their use case but are not represented in the existing set.

\subsection{The Evaluation Measurements Repository}
To support the transition from property selection to practical measurement, the cards are accompanied by an online repository \footnote{Available at \url{https://measurement-selection-app-cm6leuxsda-ew.a.run.app/}} (see the three-step guide in \Cref{fig:teaser}). The repository contains all studies included in our review from Stage 1 and can be searched by evaluation property and measurement approach. It allows researchers to examine concrete examples of measures used in prior XAI evaluations and to draw on these when designing their own studies, similar to the approach proposed by Nauta et al. \cite{Nauta2023FromAI}. Whereas existing repositories tend to emphasise objective evaluation measures, our repository adopts a broader human-centred perspective by also including subjective, user-centred, experiential, and interaction-related measures.

\subsection{Intended Usage and Project Design Stages}
\begin{figure}[!ht]
    \centering
    \Description{Diagram showing three broad design stages—Planning, Development, and Implementation—mapped onto an iterative design process. The Planning stage contains Empathise and Define; the Development stage contains Ideate and Prototype; and the Implementation stage contains Test and Implement. The main process is shown as a sequence from Empathise to Define, Ideate, Prototype, Test, and Implement. Each activity also includes an iteration loop, indicating that the process is not strictly linear. Additional feedback arrows show how later stages can lead back to earlier activities. Testing can lead back to Empathise to learn more about users, back to Ideate to find new ideas, or back to Define to redefine the problem. Iteration between Ideate and Prototype can also spark new ideas. After implementation, continued monitoring can lead back to testing. Overall, the figure illustrates that although the design process is organised into three broad stages, movement between activities and stages remains iterative.}
    \includegraphics[width=\linewidth]{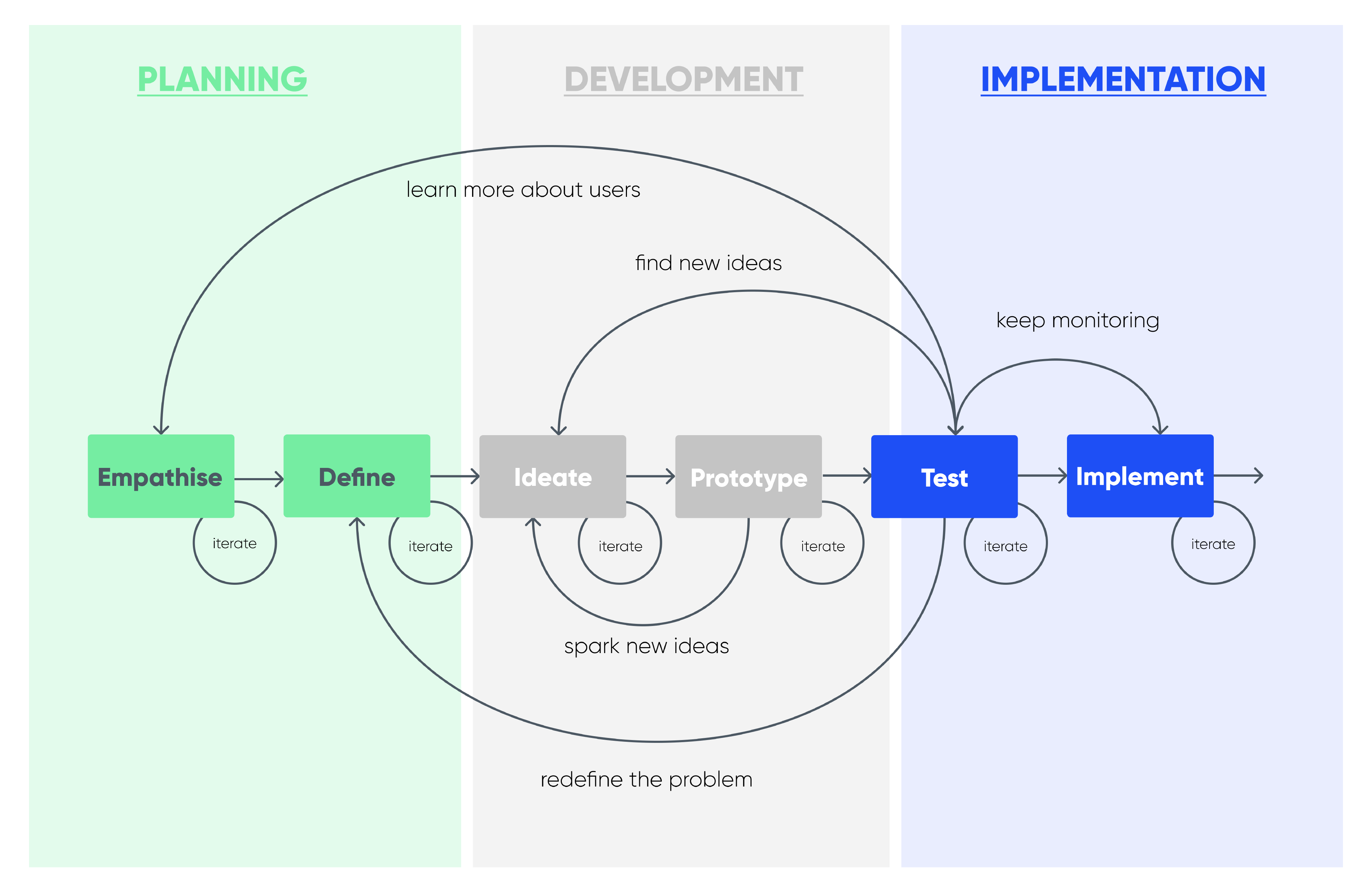}
    \caption{Design stages that show the process of user’s involvement during creating an AI system adapted from \citet{GibbonsS.2016Design101} and \citet{Dam2022WhatPopular}}
    \label{fig:Design stages}
\end{figure}
The cards are designed to support a three-step evaluation design process: \textbf{(1)} identifying what is relevant to evaluate within the research team; \textbf{(2)} prioritising what should be evaluated at the current stage of the project; and \textbf{(3)} exploring how the prioritised properties can be operationalised across different study designs and disciplinary contexts.

Before starting this process, the research team should define a clear XAI use case to which the evaluation will be applied and consider the project's current stage of development. This context helps teams assess the relevance and priority of different evaluation properties. For example, they may ask: \textit{Have users already been involved? What do we currently know about them? Is there a prototype available for evaluation?} Making these considerations explicit helps ensure that the selected properties are aligned with both the use case and the current evaluation needs of the project. 

To categorise the use cases according to their \textbf{current design stage of development}, we considered the work completed to date for each project and mapped it onto an adapted design-process model. We based this on the six phases described by the Nielsen Norman Group \cite{GibbonsS.2016Design101}—\textit{empathise, define, ideate, prototype, test,} and \textit{implement}—together with elements of the Interaction Design Foundation model \cite{Dam2022WhatPopular}. To simplify the classification, we grouped these phases into three broader stages: \textbf{planning} (\textit{empathise} and \textit{define}), \textbf{development} (\textit{ideate} and \textit{prototype}), and \textbf{implementation} (\textit{test} and \textit{implement}) (see \Cref{fig:Design stages}).

These stages reflect different evaluation needs. During the \textbf{planning} and \textbf{development} stages, researchers can still shape the type, content, and presentation of explanations and use evaluation to inform design decisions. In the \textbf{implementation} stage, evaluation is more likely to focus on assessing an existing or more mature system in use. 

It may also be beneficial for the research team to familiarise themselves with the framework and cards before the session, or to assign this responsibility to one team member who can guide the process, respond to specific questions when needed, and help the session run more smoothly. In our testing, this role was fulfilled by a moderator, who is also the first author of this paper.

\section{User Evaluation of the Method}\label{sec:study}
After developing the XAI Evaluation Cards, we subsequently tested them with research teams working on different real-world XAI projects to understand how interdisciplinary teams navigate evaluation planning collaboratively.

\subsection{Participants and Study context}
We conducted two one-hour sessions with 13 researchers, organised into six groups representing five XAI projects (\Cref{tab:freq}). Two groups (Group 5 and 6) worked independently on the same healthcare use case. The researchers who participated are colleagues of the authors and collaborate with them on different projects, but they were not directly involved in the development of the method.

The participating projects were mapped to the design stages described in Section 3.4 to examine whether the relevance and prioritisation of XAI evaluation properties differed across stages of development.

\begin{table}
\Description{Table summarising 13 participants across two sessions, organised into six groups, together with their XAI use-case domain, design stage, and current project status. Session 1 included eight computer scientists and HCI researchers focusing on XAI, divided into four groups. Group 1 worked on planning and logistics at the implementation stage, with user interviews completed and a prototype already evaluated. Group 2 worked on mental healthcare and stress management at the development stage; user feedback had been collected and a first prototype was ready but not yet evaluated. Group 3 worked on recruitment in the HR domain at the implementation stage, with user feedback collected and the prototype evaluated. Group 4 worked in education at the development stage, having conducted co-design sessions and semi-structured user interviews, while the prototype had not yet been evaluated. Session 1 included five men and three women. Session 2 included five data scientists and mathematicians working on XAI models. Group 5 worked on healthcare diagnostics at the planning stage, with data preparation for the AI model underway, the project still at an experimental stage, and the use case defined. Group 6 consisted of three participants working on the same healthcare diagnostics use case as Group 5 and participated as an exploratory online group. Their project was also at the planning stage, with data preparation underway and the use case defined. Session 2 included two men and three women.}
 \caption{List of participants and overview of use cases 
}
\centering
\label{tab:freq}
\newsavebox{\mytablebox}
  \sbox{\mytablebox}{%

  \renewcommand{\arraystretch}{1.5}
  \begin{tabular}{l>{\raggedright\arraybackslash}p{10mm}>{\raggedright\arraybackslash}p{25mm}l>{\raggedright\arraybackslash}p{75mm}}
    \toprule
    Group \#&Gender& Use case domain&Design stage&Description of the current stage\\
    \midrule
     \multicolumn{5}{p{\linewidth}}{Session 1 
(n = 8)
computer scientists and HCI researchers focusing on XAI
\newline
Moderator: social scientist}\\ 1& Male \newline Male& Planning, logistics& Implementation&User interviews, prototype has been tested\\
  2& Female \newline Male& Mental healthcare; stress management& Development&User feedback was collected, first prototype is ready but not evaluated yet\\
       3& Female\newline Male& HR domain; recruitment& Implementation&User feedback was collected, prototype has been tested\\
     4& Female \newline Male& Education domain& Development&Co-design sessions, semi-structured user interviews, prototype has not been evaluated yet\\

  \midrule
 \multicolumn{5}{p{\linewidth}}{Session 2 (n = 5)
data scientists, and mathematicians working on XAI models
\newline
Moderator: social scientist}\\ 5& Female \newline Male& Healthcare diagnostics& Planning&Data preparation for the AI model, experimental stage, use case has been defined \\
  6*& Female**\newline Male \newline Female& Healthcare diagnostics& Planning&Data preparation for the AI model, experimental stage, use case has been defined\\

  \bottomrule
\end{tabular}
}
 \usebox{\mytablebox}

  \vspace{1ex} % Small space between table and notes

  % 3. Force the notes area to match the exact width of the table
  \begin{minipage}{\wd\mytablebox}
    \small *Exploratory online group working on the same use case as Group 5. \\
**This participant was physically present during the session and is a sociologist and the last author of the paper. The participant also joined Group 5 for the final 15 minutes to take part in the property selection and prioritisation exercise.
  \end{minipage}
\end{table}

\subsection{Study Protocol and Procedure}
\begin{figure}[!ht]
    \centering
    \Description{Photograph of an example card-sorting arrangement from one research group during the XAI Evaluation Cards session. The cards are spread across a large table and grouped by category using different coloured borders. A printed overview of the complete evaluation framework is placed near the centre as a reference. The visible cards include properties from Explanation Aspects, Objective System Aspects, Subjective System Aspects, and User Experience. Within each category, cards are arranged spatially to reflect the group’s sorting and prioritisation decisions, with some cards placed in ordered sequences and others positioned separately. The photograph illustrates how participants physically manipulated and organised the cards during the collaborative evaluation-planning exercise.}
    \includegraphics[width=\linewidth]{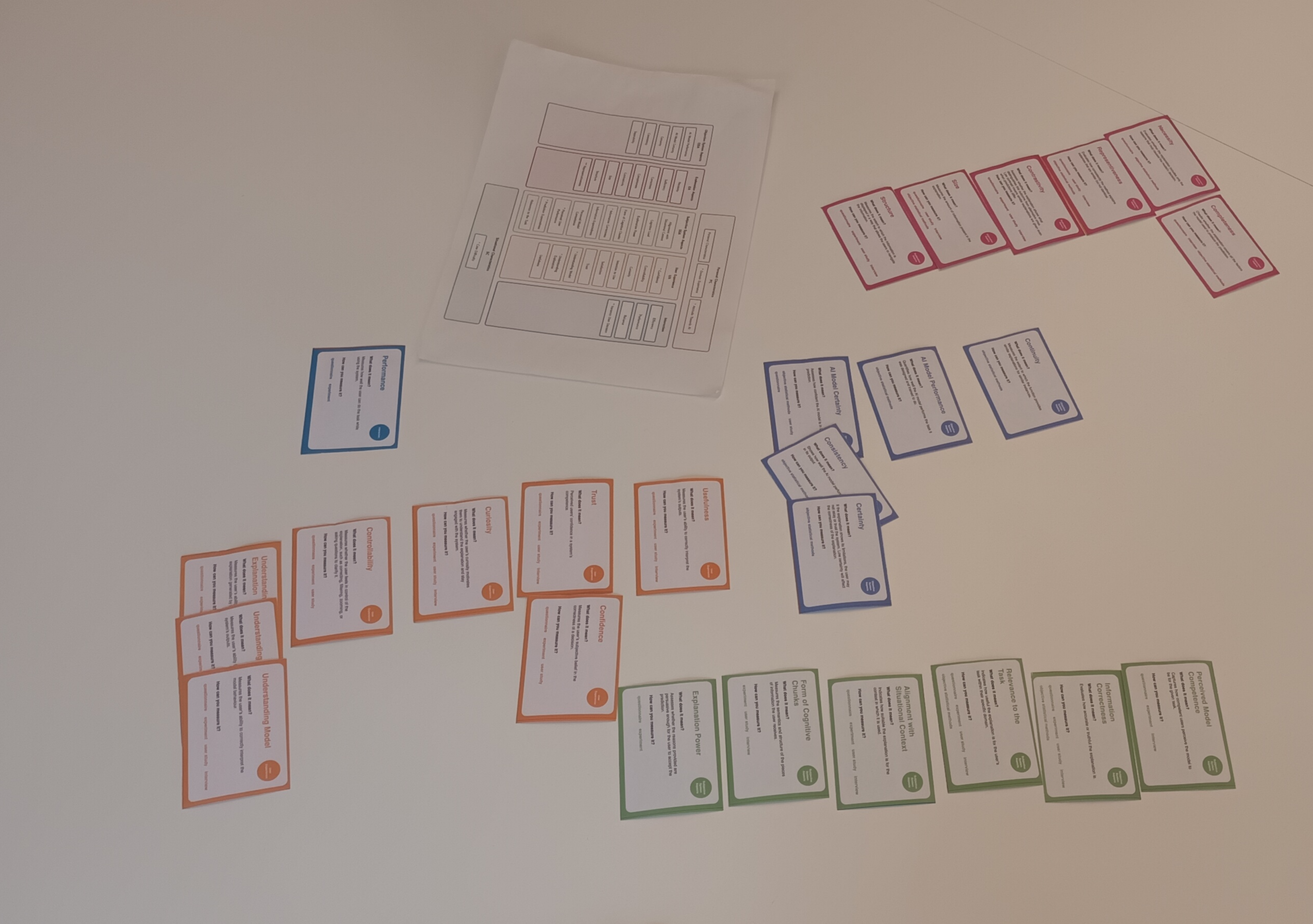}
    \caption{Example from the card sorting session}
    \label{fig:Physical cards}
\end{figure}

Each group received a complete set of XAI Evaluation Cards (see \Cref{fig:Physical cards}) and a printed version of the framework (see \Cref{fig:Framework}). The moderator, a social scientist and the first author of this paper, introduced the framework, explained how the cards were derived from it, and guided participants through the exercises. During the sessions, we intended to follow this protocol:

\begin{enumerate}
    \item \textbf{Introduction (10 min).} The moderator introduced the framework and explained the purpose and structure of the cards. Each participant introduces themselves and their XAI project they are working on and want to focus on during the session.
    \item \textbf{Forming the groups (1 min).} Participants were grouped according to the XAI projects they were working on. For sessions involving a single project, participants can instead be divided into mixed-role groups, for example comprising developers, social scientists, and designers. 
    \item \textbf{Prioritisation of categories (5 min).} Participants first ranked the five main evaluation categories according to their perceived relevance to their project. This step was primarily intended to familiarise participants with the framework and can be omitted in future sessions. 
    \item \textbf{Relevance sorting (15 min).} Participants reviewed the individual property cards and sorted them into \textit{relevant} and \textit{not relevant} for their use case. They were encouraged to discuss their reasoning and reach a shared decision. 
    \item \textbf{Prioritisation of properties (10 min).} Participants ranked the properties classified as relevant according to their importance for the evaluation, with 1 representing the highest priority. 
    \item \textbf{Next steps (5-10 min).} Throughout the exercises, the moderator checks the groups to answer clarification questions, provides guidance when needed, and takes notes on how participants discussed and negotiated the relevance of different properties. Following the sorting exercises, the moderator conducts a short discussion (5–10 min) addressing participants' experiences with the method and how they would translate the prioritisation into an evaluation plan. Questions included: \textit{How was it to discuss the properties as a group, and what made reaching consensus easier or more difficult? Did prioritising the properties help, and if so, how? What would your next steps normally be when designing the evaluation, and how would you select appropriate evaluation methods or measures?}
\end{enumerate}

\noindent\textbf{{Session 1.}}
The planned protocol was followed by all four groups. Some groups required slightly more time for the relevance-sorting and property-prioritisation exercises (Steps 4 and 5), but all groups completed the full protocol within the scheduled one-hour session. All groups prioritised properties within each of the five main categories, rather than ranking all selected relevant properties against one another. As a result, each category had its own highest-priority property or properties.

\noindent\textbf{{Session 2.}}
The procedure required several adaptations because not all participants were present in person as planned. Group 5 completed the exercise in person with the moderator, whereas Group 6 participated remotely using a Miro board with support from an additional researcher who was  online. Due to the more limited time available, the initial prioritisation of the five evaluation categories (Step 3) was omitted for both groups.

Group 5 also spontaneously combined relevance sorting and prioritisation (Steps 4 and 5). Rather than first selecting all relevant properties and subsequently prioritising them, participants ranked the properties as they determined their relevance to the use case. As in Session 1, Group 5 prioritised properties within each of the five main categories, rather than ranking all selected relevant properties against one another. Group 6 did not complete the relevance-sorting exercise within the available time and consequently did not proceed to the prioritisation step. 

Because Group 6 did not complete the core card-sorting and prioritisation activities, its card-selection data were excluded from the comparative analysis of property relevance and prioritisation. These data were incomplete and therefore not directly comparable with those of the five groups that completed the exercise. Procedural observations from Group 6 were retained only as exploratory feedback on the online implementation of the method and were not included in the comparative card-selection results.

\subsection{Data collection and Analysis}
Photographs of each group's final card selection were taken using a smartphone and transferred to a Miro board for a quick visual overview. This was then translated into Microsoft Excel by creating a dataset that recorded each group’s assigned priority for each property. We descriptively compared which properties were selected as relevant across groups and listed the frequency with which each property was selected. We also compared the priorities assigned within each group. In addition, we conducted a comparative analysis to exploratively examine the highest-priority properties (ranked 1 or 2) across the different design stages of the project.

The sessions were audio-recorded using Microsoft Teams and a smartphone voice recorder. The moderator documented notes and observations immediately after each session in a research diary, focusing on usability, points of confusion, group discussion, and opportunities for improvement. These reflections were subsequently discussed with members of the research team (the second author for Session 1; the last author for Session 2).
Audio recordings were transcribed by the Microsoft Dictate function and reviewed alongside these observations. Transcripts were used to identify illustrative quotations that contextualised the researchers' observations; no separate thematic analysis of the transcripts was conducted.
\subsection{Ethical considerations}
Participants provided oral consent prior to participation, including consent to audio recording and to the use of the collected data for research and publication. To protect participants' confidentiality and the intellectual property of ongoing projects, we report only high-level descriptions of the project domains and their design stages of development.

\section{Pilot Results}

\subsection{\textbf{Card selection and prioritisation} }

\begin{figure}[!ht]
    \centering
    \Description{Horizontal bar charts comparing the percentage of evaluation properties selected as relevant by Groups 1–5 across five categories: Objective System Aspects, Explanation Aspects, Subjective System Aspects, User Experience, and Interaction. Each panel uses the colour associated with its category. For Objective System Aspects, Groups 1–5 selected 60\%, 100\%, 40\%, 60\%, and 80\% of the available properties, respectively. For Explanation Aspects, the percentages were 75\%, 75\%, 38\%, 63\%, and 75\%. For Subjective System Aspects, they were 40\%, 90\%, 60\%, 100\%, and 60\%. For User Experience, they were 56\%, 89\%, 78\%, 56\%, and 78\%. For Interaction, they were 50\%, 100\%, 50\%, 100\%, and 25\%. Overall, the figure shows substantial variation in the breadth of properties selected across groups and categories. Group 2 selected a high proportion of properties across all five categories, while Groups 3 and 5 were more selective in some categories. Group 4 selected all Subjective System Aspects and Interaction properties, whereas Group 5 selected only 25\% of the Interaction properties.}
    \includegraphics[width=\linewidth]{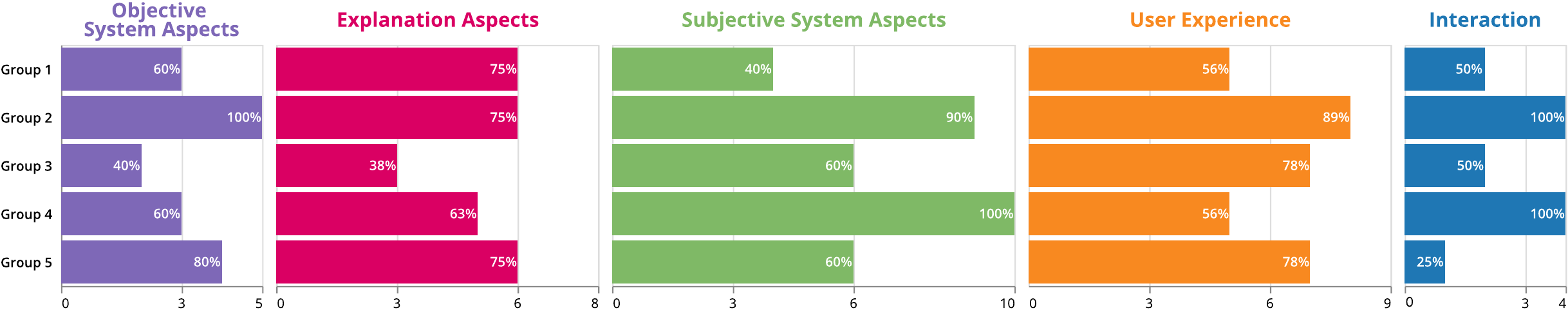}
    \caption{Overview and frequency of property selections across the five main evaluation categories}
    \label{fig:results_by_group}
\end{figure}

The number and type of properties selected as relevant varied considerably across groups. These differences appeared to relate both to the characteristics of the use case and to the stage of the project. For example, Group 2, which was still developing and evaluating its prototype, selected a broad range of properties across all five categories (see \Cref{fig:results_by_group}). In contrast, Group 3, which was further along in the project and had already conducted evaluation activities, was more selective, particularly for Objective System and Explanation Aspects.

A somewhat different pattern emerged for Group 5, whose project was at the earliest stage of development (no working prototype, only a conceptual phase). This group selected a high proportion of Objective System Aspects (80\%), such as AI Model Performance, Consistency, and Separability, but fewer Subjective System Aspects (60\%) and Interaction properties (25\%). During the discussion, several of these user-oriented properties remained relatively abstract because the system and its interaction context had not yet been fully defined. 

Despite considerable variation between groups, some common priorities emerged. AI Model Performance was ranked as the most important Objective System Aspect by three of the four groups in the first session. Other properties were much more project-specific. For example, Satisfaction and Curiosity were selected only by Group 2, while Understanding Model Behaviour was selected only by Group 3 and Completeness only by Group 2.

No property in the framework was rejected by all groups: every property was considered relevant by at least one group. This provides initial support for maintaining a broad set of evaluation properties while allowing research teams to select and prioritise those that are relevant to their particular use case.

\begin{figure}[!ht]
    \centering
    \Description{Dot plot showing the highest-priority evaluation properties, defined as properties ranked first or second by participating groups, across three design stages: planning, development, and implementation. The horizontal axis lists individual evaluation properties, while the vertical axis represents the three design stages. Each dot indicates that a property was ranked among the two highest priorities by at least one group at that stage. Dot colours distinguish seven categories: Personal Characteristics, Situational Characteristics, Objective System Aspects, Explanation Aspects, Subjective System Aspects, User Experience, and Interaction. The figure shows that high-priority properties are distributed across multiple categories and that their selection varies by design stage. Planning displays priorities across a particularly broad range of properties, while development and implementation show different, partly overlapping patterns. Overall, the visualisation illustrates that the evaluation properties considered most important are not identical across design stages.}
    \includegraphics[width=\linewidth]{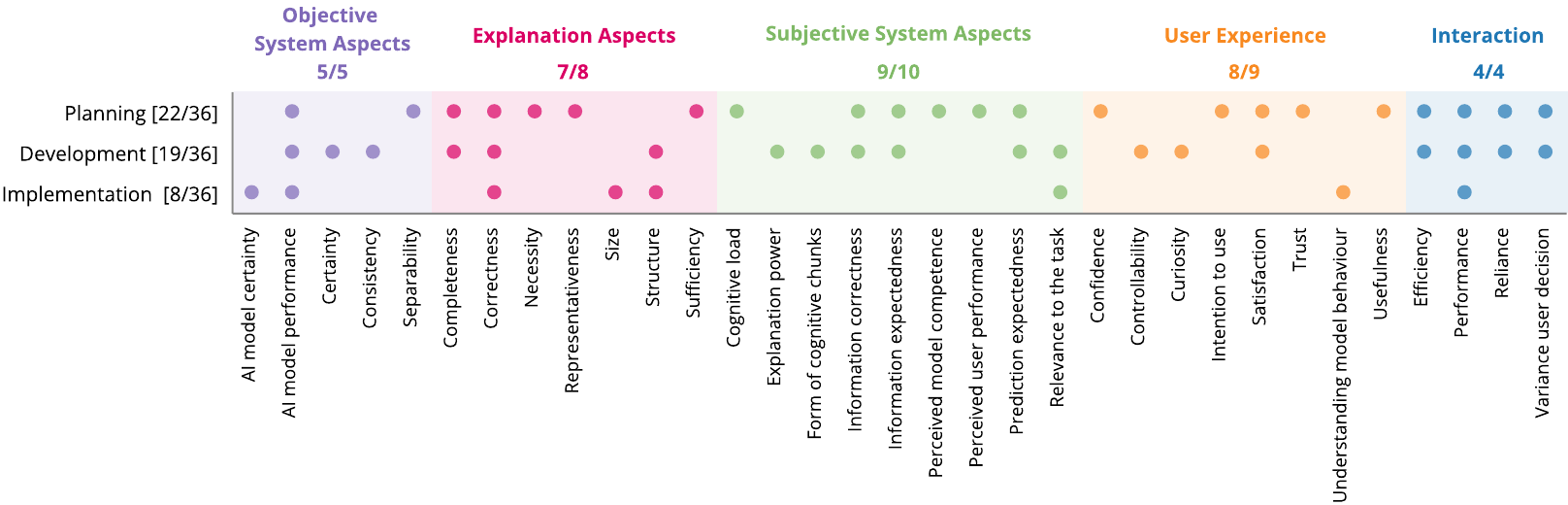}
    \caption{Highest-priority evaluation properties across design stages, showing only properties ranked first or second by each group}
    \label{fig:Design stages and priorities}
\end{figure}

As we can see in \Cref{fig:Design stages and priorities}, the most important properties varied across design stages. Among the groups in the implementation stage, the highest-ranked properties formed a relatively concentrated set, with notable overlap between groups despite working on different use cases. This may indicate that, once a system is more mature and evaluation goals are more clearly defined, teams converge on a narrower set of properties they consider most important. The development-stage groups selected a somewhat broader range of high-priority properties, although some common priorities remained visible. For the planning stage, only one group was available, preventing comparison across groups; however, its priorities were more widely distributed across the framework. This broader selection may reflect the greater uncertainty at an early stage, when key aspects of the use case, users, and system design are still being defined. Given the small number of groups, these patterns should be interpreted as exploratory rather than as evidence of systematic differences between development stages.

 \subsection{\textbf{How can the card sorting support XAI evaluation} }
 Participants described their existing process for selecting evaluation properties and measures as strongly dependent on searching and comparing previous studies. Several participants explained that they typically start from the literature, identify constructs and measures used in related studies, and adapt these to their own context. As one participant from Group 1 explained: “It will always have to be taking an existing framework and adapting it rather than using it as it was supposed to be.”

 Similarly, a participant from Group 3 described needing to “start looking for all these papers” when encountering a property they had not previously measured. Participants therefore valued having a structured overview of potential evaluation properties in one place. 

Participants generally considered the prioritisation step useful because evaluating all potentially relevant properties was regarded as unrealistic. One researcher summarised this tension directly: “Because you can do so many things. I want to do a lot of things. It’s impossible actually.” (participant from Group 3)

 The prioritisation exercise therefore served not only to identify relevant properties but also to distinguish between properties that were potentially interesting and those that participants considered sufficiently important to include in an actual evaluation.

 However, participants did not always prioritise at the level of individual properties. Group 1, for example, reasoned across the five broader categories and considered excluding entire categories depending on the purpose and maturity of the system. A researcher from Group 1 explained that an early prototype might require greater attention to improving explanations, whereas a more developed dashboard could shift attention towards user experience or other aspects of evaluation. This suggests that participants moved between category-level and property-level reasoning when constructing an evaluation strategy.

Across several groups, participants emphasised that identifying a relevant property did not resolve the subsequent question of how it should be measured. Their current approach was typically to search the literature for validated questionnaires or measures associated with the selected construct. A participant from Group 1 described this as “the hardest part,” particularly when existing measures did not fully fit the use case. This could be tackled by offering the research team a repository with possibilities of concrete measurements.

 Participants also wanted information about which properties are commonly evaluated together. Groups 1, 2, and 5 raised this explicitly. For example, a participant from Group 2 explained: \say{If you're looking for a specific construct and then you want to find the validated questionnaire, then you see that it's usually associated with other constructs, like satisfaction and usefulness; they're often measured together.}.

When we asked about the next step, which would be the repository of methods. Participants suggested that the supporting repository should not only list individual measures, but also provide information about relationships between properties, validated instruments, and the types of studies in which particular measures have previously been used. One participant from Group 4 proposed indicating whether a method is suited to quantitative, qualitative, or mixed-method evaluations and whether it requires large or small participant samples. Nevertheless, majority of the groups expressed their interest in repository and found it as potentially very useful tool.

The sessions also revealed limitations in using the framework without additional support. Participants occasionally struggled with property definitions, particularly when they were familiar with the same or closely related concepts under different terminology (for example: difference between completeness and sufficiency). In some cases, participants questioned whether two properties represented sufficiently distinct concepts or should instead be combined.

 The moderator played an important role in resolving these ambiguities (e.g. by giving the participants a concrete example how this property can be applied) and reminding participants to judge properties in relation to their specific use case. Participants also responded positively to concrete examples illustrating how individual properties could apply in practice (remark from Group 5). These observations suggest that future versions of the cards could benefit from more recognisable real-life examples and clearer distinctions between related properties.

 A further indication of the importance of facilitation emerged from the exploratory online session. Participants in this session did not complete the full exercise and encountered similar difficulties with terminology. This may have been influenced by several factors, including limited familiarity with the framework and cards, a less clearly defined use case due to the early stage of the project, and the online format itself. 

\section{Discussion}

The two evaluation sessions suggest that the XAI Evaluation Cards can provide a shared and structured space for evaluation planning. By making a broad range of evaluation properties visible at the same time, the cards helped participants consider options beyond those they would normally retrieve from familiar literature or previously used measures. This supported discussion about which properties were relevant to a particular use case and which should be prioritised at the current stage of development, which is consistent with previous calls for combining both objective and subjective measurements \cite{Mohseni2021ASystems, Schmid2022WhatFar}.

At the same time, the sessions showed that evaluation priorities are highly context-dependent. Our findings extend previous work on the ranking of XAI evaluation criteria \cite {Liao2022ConnectingAI} indicating that priorities may need to be determined in relation to the specific use case and current design stage. Participants' selections differed according to the use case, design stage, and disciplinary background of the researchers involved. This suggests the idea that XAI evaluation should not be approached through a universal checklist of properties. Instead, the cards are intended to structure the process through which teams decide what is relevant for their specific context and design stages. Based on our observations, we identified several practical considerations for applying the method.

\subsection{Facilitation supports the use of the cards}

Having a moderator who was familiar with the framework was particularly useful when participants encountered unclear or overlapping definitions. The moderator also helped participants remain focused on their specific use case and encouraged them to distinguish between properties that may be generally important and those that were relevant to evaluate at the current stage of their project. This suggests that facilitation may be particularly valuable when teams use the method for the first time. Future iterations of the cards could reduce this dependency by providing clearer definitions and more concrete examples. The need for clarification during the sessions reflects the terminology problems previously identified in XAI research \cite{Graziani2023ASciences, Clinciu2019ATerminology}. For the current version, we recommend that at least one member of the research team becomes familiar with the framework and cards beforehand and can support the process when needed.

\subsection{Multidisciplinary participation broadens evaluation planning}

The sessions also suggest that involving researchers with different disciplinary backgrounds can broaden the discussion around evaluation priorities. Although reaching consensus may require additional time, contrasting perspectives encouraged participants to articulate why particular properties mattered and to consider aspects they might otherwise have overlooked \cite{Bennett2026}. This was particularly visible in Session 2, where researchers with a social science background introduced additional user-centred considerations into a discussion that otherwise focused strongly on technical aspects of the system.

The value of the method may therefore lie not only in the final selection of cards, but also in the discussion through which this selection is reached. In this sense, the cards function as a shared vocabulary that can help interdisciplinary teams surface assumptions, negotiate priorities, and establish a common understanding of what should be evaluated.

\subsection{Evaluation priorities change across design stages of development}

The relevance of individual properties appeared to vary according to project maturity, design stage, and the evaluation activities that had already taken place. Projects at an earlier stage sometimes prioritised technical or system-level properties because user interaction was not yet sufficiently defined, whereas projects further along in development could make more targeted decisions informed by previous evaluation activities. In our sample, projects in the implementation stage also showed a more selective pattern of prioritisation, with a narrower set of properties ranked as highest priority than in the earlier design stages.

This suggests that the card-sorting exercise should not necessarily be treated as a one-time activity. Research teams could revisit the cards at different stages of development, as the properties considered relevant to evaluate may change as the system, use case, and available evidence evolve. The method therefore supports evaluation planning as an iterative rather than static process, as illustrated in \Cref{fig:Design stages}.

\subsection{The interaction format should support collaborative reflection}

Participants engaged actively with the physical cards, rearranging them throughout the discussion as their priorities evolved. The exploratory online session was less successful, but several factors differed simultaneously, including the level of facilitation, familiarity with the use case, and interaction format. We therefore cannot conclude that physical cards are inherently superior to a digital version.

Instead, these findings suggest that a future digital implementation should preserve the flexibility and collaborative character of the physical exercise, allowing researchers to compare and reorganise properties while discussing their choices with others. The design challenge is therefore not simply to digitise the cards, but to retain the collaborative and reflective qualities associated with card-based design methods \cite {Aarts2020DesignSets} that are currently supported by the physical XAI Evaluation Cards exercise.

\subsection{From identifying properties to selecting measurements}

The sessions further showed that identifying relevant properties is only one step in XAI evaluation planning. Participants repeatedly raised questions about how selected properties should subsequently be operationalised and which validated measures or methodological approaches were available. Several participants also expressed interest in seeing which properties are commonly evaluated together.

This reinforces the distinction underlying our method between deciding \textit{what} should be evaluated and deciding \textit{how} it should be measured. The cards primarily support the former by helping teams identify and prioritise evaluation properties, while the accompanying repository supports the latter by providing examples of how these properties have previously been operationalised. Together, they provide a three-step process: identifying potentially relevant properties, prioritising them for the current use case and stage of development, and exploring appropriate ways of measuring them.

\section{Limitations and Future work}

This study represents an initial testing of the XAI Evaluation Cards and has several limitations. First, the method was tested with only six groups across a limited number of projects. Further evaluation is therefore needed across a broader range of domains, project types, team compositions, and design stages of development to assess how consistently the method supports evaluation planning.

Second, the study relied on a convenience sample of researchers who are colleagues of the authors and collaborate on shared projects. While this allowed us to observe real-world, cross-functional teams, it introduces the potential for social desirability bias, as colleagues might be more inclined to view the tool positively or align with the proposed framework. Furthermore, Groups 1-4 were already somewhat familiar with the underlying framework and regularly worked with several of the evaluation properties included on the cards. This familiarity may have made it easier for them to interpret and prioritise the properties compared with Groups 5 and 6, who had less prior experience. However, because this study was designed as an early-stage formative pilot to observe the mechanism of boundary negotiation rather than a summative validation of the toolkit's ultimate efficacy, these sessions still provided critical insight into how the cards facilitate interdisciplinary discussion. Future studies should nevertheless include independent teams with varying levels of XAI experience to assess the method's broader usability and mitigate these biases. 

Third, some participants experienced difficulties interpreting particular property definitions or recognised the same concepts under different terminology. Future iterations should therefore include clearer definitions, and concrete examples drawn from domains that are accessible to researchers from different disciplinary backgrounds. This may also reduce the reliance on moderator support during the exercise.

The accompanying repository remains under development. We aim to extend it with more concrete examples of validated measures, information about how properties are commonly operationalised together, and guidance on which methods may be appropriate for different study designs and contexts. Further user testing of the repository is also needed to better understand the needs of XAI researchers and to refine its structure, content, and usability accordingly.

Finally, the current study focused primarily on how researchers use the cards to discuss and prioritise evaluation properties. It does not yet demonstrate whether using the method leads to more appropriate, comprehensive, or effective XAI evaluations in practice. Future work should therefore examine how the selected properties are translated into concrete study designs and whether the method influences subsequent evaluation decisions over time. We are also exploring the use of the cards with end users to investigate whether the method can support the elicitation and comparison of evaluation priorities beyond research teams. For these audiences, further adaptation may be necessary, including less abstract terminology and more concrete, accessible examples tailored to the specific usage context.

\section{Conclusion}
In this paper, we introduced and initially tested the XAI Evaluation Cards, an evidence-informed method designed to support reflective and collaborative XAI evaluation planning. The cards provide research teams with a shared design space in which they can explore potentially relevant evaluation properties, discuss their relevance to a specific use case and design stage, and prioritise them with guidance on general measurement approaches that could subsequently be used. Our initial testing suggests that research teams found the exercise helpful for structuring these discussions and welcomed the idea of an accompanying human-centred XAI evaluation repository to support the translation of prioritised properties into concrete measures and study designs.

The findings further suggest that the cards can broaden the range of evaluation properties considered beyond those researchers might otherwise select based on familiarity or disciplinary background. By bringing together objective and subjective properties and making their relationships visible, the method encourages research teams to consider XAI evaluation more comprehensively while still making context-dependent choices about what is relevant to evaluate. Rather than prescribing a fixed set of criteria, the XAI Evaluation Cards support a situated and iterative evaluation planning process in which priorities can be revisited as the system, use case, and available evidence evolve.

\bibliography{references_K, references_I, sample-base}
\bibliographystyle{ACM-Reference-Format}

\clearpage
\appendix



\section*{Supplementary material}

## Explanation Aspects

properties evaluating the quality and effectiveness of the system's explanations

## Explanation Aspects

properties evaluating the quality and effectiveness of the system's explanations

## Explanation Aspects

properties evaluating the quality and effectiveness of the system's explanations

## Subjective System Aspects

properties assessing users' perceptions of the objective system aspects

## User Experience (UX)

properties assessing what users perceive, feel, and experience when interacting with the system

## Objective System Aspects

properties specific to the AI system as a whole, with a main focus on its technical capabilities

## Subjective System Aspects

properties assessing users' perceptions of the objective system aspects

## User Experience (UX)

properties assessing what users perceive, feel, and experience when interacting with the system

## Interaction

properties measuring the actual interaction between the user and the system's explanations

## Structure

Explanation Aspects

**What does it mean?**

Measures the extent to which the information is displayed in a way that allows the users to navigate the information.

**How can you measure it?**

questionnaire experiment user study interview

## Representiveness

Explanation Aspects

**What does it mean?**

Evaluates how consistently the method explains instances that are similar but not identical.

**How can you measure it?**

questionnaire user study interview

objective statistical methods

Explanation Aspects

**What does it mean?**

**How can you measure it?**

Objective System Aspects

**What does it mean?**

**How can you measure it?**

User Experience

**What does it mean?**

**How can you measure it?**

Subjective System Aspects

**What does it mean?**

**How can you measure it?**

Interaction

**What does it mean?**

**How can you measure it?**

User Experience

**What does it mean?**

**How can you measure it?**

Subjective System Aspects

**What does it mean?**

**How can you measure it?**

## Subjective System Aspects

properties assessing users' perceptions of the objective system aspects

## Subjective System Aspects

properties assessing users' perceptions of the objective system aspects

## Subjective System Aspects

properties assessing users' perceptions of the objective system aspects

## Explanation Aspects

properties evaluating the quality and effectiveness of the system's explanations

## Explanation Aspects

properties evaluating the quality and effectiveness of the system's explanations

## Explanation Aspects

properties evaluating the quality and effectiveness of the system's explanations

## Explanation Aspects

properties evaluating the quality and effectiveness of the system's explanations

## Explanation Aspects

properties evaluating the quality and effectiveness of the system's explanations

## Explanation Aspects

properties evaluating the quality and effectiveness of the system's explanations

## Perceived Model Competence

Subjective System Aspects

**What does it mean?**

Captures how competent users perceive the model to be for the given task.

**How can you measure it?**

questionnaire experiment interview

## Relevance to the Task

Subjective System Aspects

**What does it mean?**

Indicates how useful the explanation is for the user's task within their specific domain.

**How can you measure it?**

questionnaire experiment user study interview objective statistical methods

## Perceived User Performance

Subjective System Aspects

**What does it mean?**

Captures how competent users feel they themselves are when performing the task.

**How can you measure it?**

questionnaire user study interview

## Necessity

Explanation Aspects

**What does it mean?**

Evaluates whether the explanation includes only the reasons that truly caused the model's decision.

**How can you measure it?**

questionnaire objective statistical methods

## Completeness

Explanation Aspects

**What does it mean?**

Checks whether the explanation covers all the factors the model used to produce the prediction.

**How can you measure it?**

questionnaire interview objective statistical methods

## Sufficiency

Explanation Aspects

**What does it mean?**

Assesses whether the explanation includes all the reasons behind the decision and whether these reasons are sufficient without requiring additional information.

**How can you measure it?**

questionnaire user study interview

## Contrastivity

Explanation Aspects

**What does it mean?**

Assesses whether the explanation shows what distinguishes this outcome from alternative ones. Low contrast means similar explanations are given even when predictions differ.

**How can you measure it?**

questionnaire experiment user study interview

## Correctness

Explanation Aspects

**What does it mean?**

Evaluates how closely the chosen reasons match the model's actual internal reasoning.

**How can you measure it?**

objective statistical methods user study interview

## Size

Explanation Aspects

**What does it mean?**

Measures the amount of information present in the explanation.

**How can you measure it?**

questionnaire user study interview objective statistical methods

## Interaction

properties measuring the actual interaction between the user and the system’s explanations



## Subjective System Aspects

properties assessing users’ perceptions of the objective system aspects

## Alignment with Situational Context

Subjective System Aspects

**What does it mean?**

Indicates how suitable the explanation is for the context in which it is used.

**How can you measure it?**

questionnaire experiment user study interview

## Reliance

Interaction

**What does it mean?**

Measures whether the user is willing to provide control to the machine for the given task.

**How can you measure it?**

questionnaire user study interview

## Variance User Decision

Interaction

**What does it mean?**

Measures users' agreement in a decision-making scenario. Is the outcome what I expected?

**How can you measure it?**

questionnaire experiment

## Cognitive load

Subjective System Aspects

**What does it mean?**

Measures the cognitive effort the user has to do to achieve the task.

**How can you measure it?**

questionnaire experiment user study interview

## Explanation Power

Subjective System Aspects

**What does it mean?**

Assesses whether the reasons provided are persuasive enough for the user to accept the prediction.

**How can you measure it?**

questionnaire experiment

## Form of Cognitive Chunks

Subjective System Aspects

**What does it mean?**

Measures the semantics and structure of the pieces of information the user receives.

**How can you measure it?**

experiment user study interview

## Information Correctness

Subjective System Aspects

**What does it mean?**

Evaluates how accurate or truthful the explanation is.

**How can you measure it?**

questionnaire experiment user study interview

objective statistical methods

## Information Expectedness

Subjective System Aspects

**What does it mean?**

Measures how much information presented in the explanation was anticipated by the user.

**How can you measure it?**

questionnaire experiment user study interview

## Prediction Expectedness

Subjective System Aspects

**What does it mean?**

Measures how much the AI model prediction was anticipated by the user.

**How can you measure it?**

questionnaire experiment user study interview

## Confidence

User Experience

**What does it mean?**

Measures the user's subjective belief in the correctness of a decision.

**How can you measure it?**

questionnaire experiment user study

## Controllability

User Experience

**What does it mean?**

Measures whether the user feels in control of the explanation, such as correcting, filtering, zooming, or asking questions to clarify it.

**How can you measure it?**

questionnaire experiment user study

## Curiosity

User Experience

**What does it mean?**

Measures whether the user's curiosity motivates them to understand the explanation and stay engaged with the system.

**How can you measure it?**

questionnaire experiment user study interview

## Intention to Use

User Experience

**What does it mean?**

Measures an individual's willingness to use an explanation in the future.

**How can you measure it?**

questionnaire user study interview

## Satisfaction

User Experience

**What does it mean?**

Measures the level of fulfilment the user gets while interacting with the explanation.

**How can you measure it?**

questionnaire user study interview

## Understanding

User Experience

**What does it mean?**

Measures the user's ability to correctly interpret the system's outputs.

**How can you measure it?**

questionnaire experiment user study interview

## Performance

Interaction

**What does it mean?**

Measures how well the user can do the task while using the system.

**How can you measure it?**

questionnaire experiment

## Efficiency

Interaction

**What does it mean?**

Measures the speed at which a task can be performed.

**How can you measure it?**

questionnaire experiment

## Understanding Explanation

User Experience

**What does it mean?**

Measures the user's ability to correctly interpret the explanation generated by the system.

**How can you measure it?**

questionnaire experiment user study interview

## User Experience (UX)

properties assessing what users perceive, feel, and experience when interacting with the system

## User Experience (UX)

properties assessing what users perceive, feel, and experience when interacting with the system

## User Experience (UX)

properties assessing what users perceive, feel, and experience when interacting with the system

## User Experience (UX)

properties assessing what users perceive, feel, and experience when interacting with the system

## User Experience (UX)

properties assessing what users perceive, feel, and experience when interacting with the system

## User Experience (UX)

properties assessing what users perceive, feel, and experience when interacting with the system

## User Experience (UX)

properties assessing what users perceive, feel, and experience when interacting with the system

## Interaction

properties measuring the actual interaction between the user and the system’s explanations

## Interaction

properties measuring the actual interaction between the user and the system’s explanations

## Objective System Aspects

properties specific to the AI system as a whole, with a main focus on its technical capabilities

## Objective System Aspects

properties specific to the AI system as a whole, with a main focus on its technical capabilities

## Objective System Aspects

properties specific to the AI system as a whole, with a main focus on its technical capabilities

## Objective System Aspects

properties specific to the AI system as a whole, with a main focus on its technical capabilities

## Objective System Aspects

properties specific to the AI system as a whole, with a main focus on its technical capabilities

## Objective System Aspects

properties specific to the AI system as a whole, with a main focus on its technical capabilities

## User Experience (UX)

properties assessing what users perceive, feel, and experience when interacting with the system

## User Experience (UX)

properties assessing what users perceive, feel, and experience when interacting with the system

## User Experience (UX)

properties assessing what users perceive, feel, and experience when interacting with the system

## AI Model Performance

Objective System Aspects

**What does it mean?**

Quantifies how well the AI model performs the task it was designed and trained to do.

**How can you measure it?**

objective statistical methods

## Consistency

Objective System Aspects

**What does it mean?**

Shows how well the AI model performs is consistent in its output.

**How can you measure it?**

objective statistical methods

## Separability

Objective System Aspects

**What does it mean?**

The XAI method should provide different explanations for different instances.

**How can you measure it?**

objective statistical methods

## AI Model Certainty

Objective System Aspects

**What does it mean?**

Assesses how confident the AI model is in its own prediction.

**How can you measure it?**

objective statistical methods · user study · questionnaire

## Certainty

Objective System Aspects

**What does it mean?**

If the explanation shows its limitations, the user may not relay or trust the system. Low certainty will affect the correctness of the explanation.

**How can you measure it?**

objective statistical methods

## Continuity

Objective System Aspects

**What does it mean?**

Measures the extent to which the function provides similar explanations for similar instances.

**How can you measure it?**

objective statistical methods

## Trust

User Experience

**What does it mean?**

Perceived users' confidence in a system's competence.

**How can you measure it?**

questionnaire · experiment · user study · interview

## Usefulness

User Experience

**What does it mean?**

Measures the user's ability to correctly interpret the system's outputs.

**How can you measure it?**

questionnaire · experiment · user study · interview

## Understanding Model

User Experience

**What does it mean?**

Measures the user's ability to correctly interpret the model behaviour.

**How can you measure it?**

questionnaire · experiment · user study · interview



\end{document}